\documentclass[%
superscriptaddress,
reprint,
bibnotes,
 amsmath,
 amssymb,
 aps,
pra,
]{revtex4-1}

\usepackage{amsmath}  
\usepackage{amsfonts} 
\usepackage{graphicx} 
\usepackage{enumitem}
\usepackage{xcolor}
\usepackage{appendix}
\usepackage{multirow}

\begin{document}


\title{Design, Analysis, Tools, and Apprenticeship (DATA) Lab}

\author{Kelsey Funkhouser}
\email{kfunkh@msu.edu} 
\affiliation{Department of Physics \& Astronomy, Michigan State University, East Lansing, MI, 48824}
\author{William M. Martinez}
\affiliation{Vanderbilt Institute of Nanoscale Science and Engineering, Vanderbilt University, Nashville, TN 37235}
\author{Rachel Henderson}
\affiliation{Department of Physics \& Astronomy, Michigan State University, East Lansing, MI, 48824}
\author{Marcos D. Caballero}
\email{caballero@pa.msu.edu}
\affiliation{Department of Physics \& Astronomy, Michigan State University, East Lansing, MI, 48824}
\affiliation{CREATE for STEM Institute, Michigan State University, East Lansing, MI, 48824}
\affiliation{Department of Physics \& Center for Computing in Science Education, University of Oslo, N-0316 Oslo, Norway}


\date{\today}

\begin{abstract}
Recently, there have been several national calls to emphasize physics practices and skills within laboratory courses. In this paper, we describe the redesign and implementation of a two-course sequence of algebra-based physics laboratories at Michigan State University called Design Analysis Tools and Apprenticeship (DATA) Lab. The large-scale course transformation removes physics specific content from the overall learning goals of the course, and instead, uses physics concepts to focus on specific laboratory practices and research skills that students can take into their future careers. Students in DATA Lab engage in the exploration of physical systems to increase their understanding of experimental process, data analysis, collaboration, and scientific communication. In order to ensure our students are making progress toward the skills outlined in the course learning goals, we designed all of the assessments in the courses to evaluate their progress specific to these laboratory practices. Here, we will describe the structures, scaffolds, goals, and assessments of the course.
\end{abstract}

\maketitle 

\section{Introduction}\label{sec:intro}

New knowledge in physics is driven by the observation of phenomena, the design of experiments to probe these phenomena, and the communication of and debate around the resulting measurements in public fora. Laboratory courses in physics are thus unique spaces where students can engage in these central aspects of studying physical systems. Greater emphasis on these aspects in laboratory spaces is needed to accurately represent the physics discipline and to engage students in the universal scientific endeavor that is driven by observation, measurement, and communication.

Recently, national calls have been made to design laboratory instruction such that it emphasizes students' engagement in experimental scientific practices rather than simply re-enforcing content learning \cite{kozminski2014aapt,holmes2017value}. Such experiences would be better aligned with discovery-based learning \cite{olson2012engage}, which is more representative of the enterprise of experimental physics. This focus on science practices is articulated in the American Association of Physics Teachers' {\it Recommendations for the Undergraduate Physics Laboratory Curriculum} \cite{kozminski2014aapt}. These recommendations call for all laboratories in undergraduate physics to better represent experimental physics by constructing laboratory curriculum around science practices such as designing experiments, analyzing and visualizing data, and communicating physics. Arguably, middle-division and advanced laboratory courses for physics and astronomy majors -- with their more complex experiments and equipment as well as their focus on the professional development of future physicists -- tend to engage students with these practices.

By contrast, introductory physics laboratory courses tend to have more prescriptive and direct approaches to instruction. In these courses, students often follow a well-documented procedure and do not typically have opportunities to explore the observed phenomenon and the associated experimental work. At larger universities in the United States, these introductory laboratory courses are taught to thousands of students per semester, which makes these more direct approaches to instruction attractive as they are quite efficient. At many US schools, engineering students, physical science majors, and biological science students must pass these laboratory courses to complete their degree program. The scale of these course offerings provides an additional challenge to incorporating science practices. There are unique examples in the literature where students of introductory physics are engaged with scientific practices such as the Investigative Science Learning Environment (ISLE) \cite{etkina2001investigative} and Studio Physics \cite{wilson1994cuple}. However, these courses have the advantage of being taught to smaller population of students than most introductory laboratory courses, in the case of ISLE, or having an integrated ``lecture'' and a modified instructional space, in the case of Studio Physics, and thus can make use of greater instructional resources.

In this paper, we describe a stand-alone, introductory physics laboratory course sequence for biological science majors at Michigan State University (MSU) that was designed specifically to engage students in scientific practices through the work of experimental physics. Students learn to design experiments, analyze and visualize their data, and communicate their results to their peers and instructors. Design, Analysis, Tools, and Apprenticeship (DATA) Lab is unique in that it is was explicitly designed with the AAPT Lab Recommendations in mind. The sequence is a stand-alone mechanics laboratory (DL1) and a separate E\&M and optics laboratory (DL2), which is taught to more than 2000 students per year. Furthermore, the process of developing and launching this pair of courses required that we confront and overcome several well-documented challenges such as
departmental norms for the course, expectations of content coverage, and the lack of instructor time \cite{dancy2008barriers}.

\begin{table*}[th]
\caption{Finalized learning goals for DATA Lab}\label{tab:lgs}
\begin{tabular}{l|p{4in}}
\hline
\hline
Learning Goal & Description \\
\hline
     LG1 - Experimental Process&  Planning and executing an experiment to effectively explore how different parameters of a physical system interact with each other. Generally taking the form of model evaluation or determination.\\
     LG2 - Data Analysis & Knowing how to turn raw data into an interpretable result (through plots, calculations, error analysis, comparison to an expectation, etc.) that can be connected to the bigger physics concepts.\\
     LG3 - Collaboration & Working effectively as a group. Communicating your ideas and understanding. Coming to a consensus and making decisions as a group.\\
     LG4 - Communication & Communicating understanding -- of the physics, the experimental process, the results -- in a variety of authentic ways -- to your peers, in a lab notebook, in a presentation or proposal. \\
     \hline
\end{tabular}
\end{table*}

We begin this paper by describing how the learning goals for the lab sequence were constructed through a consensus-driven process (Sec.~\ref{LGSection}). In Sec.~\ref{structures}, we provide an overview of the course structure -- diving deeper into the details of the course materials later (Sec.~\ref{OverviewSection}). We describe the assessments for this course in Sec.~\ref{assessments} as they are somewhat non-traditional for a course of this level and scale. To make our discussion concrete, we highlight a particular example in Sec.~\ref{ExpOverSection}. Finally, we offer a measure of efficacy using student responses to the Colorado Learning Attitudes about Science Survey for Experimental Physics \cite{zwickl2014epistemology} (Sec.~\ref{efficacy}) and some concluding remarks (Sec.~\ref{conclusions})

\section{Learning Goals}\label{LGSection}

As this laboratory course serves the largest population of students enrolled in introductory physics at MSU, it was critical to develop a transformed course that reflected faculty voice in the design. While physics faculty are not often steeped in formal aspects of curriculum development, sustained efforts to transform physics courses take an approach where faculty are engaged in the process to develop a consensus design \cite{chasteen2011thoughtful,chasteen2012transforming,wieman2017improving}. In this process, interested faculty are invited to participate in discussions around curriculum design, but experts in curriculum and instruction synthesize those discussions to develop course structures, materials, and pedagogy. These efforts are then reflected out to faculty to iterate on the process. Our design process followed the approach developed by the University of Colorado's Science Education Initiative \cite{chasteen2011thoughtful,chasteen2012transforming,wieman2017improving}. In this process, faculty are engaged in broad discussions about learning goals, the necessary evidence to achieve the expected learning, and the teaching practices and course activities that provide evidence that students are meeting these goals. Below, we discuss the approach to developing learning goals for the course as well as present the finalized set of learning goals from which the course was designed. We refer readers to \citet{wieman2017improving} for a comprehensive discussion of setting about transforming courses at this scale.

Prior to engaging in curriculum and pedagogical design, an interview protocol was developed to talk with faculty about what they wanted students to get out of this laboratory course once students had completed the two semester sequence. The interview focused discussion on what made an introductory laboratory course in physics important for these students and what role it should play as a distinct course since, at MSU, students do not need to enroll in the laboratory course at the same time as the associated lecture course. A wide variety of faculty members were interviewed including those who had previously taught the course, those who had taught other physics laboratory courses, and those who conduct experimental research. In total, 15 interviews were conducted with faculty. This number represents more than half of the total number of experimental faculty who teach at MSU.

The discussion of faculty learning goals was wide-ranging and covered a variety of important aspects of laboratory work including many of the aspects highlighted in the AAPT Laboratory Guidelines \cite{kozminski2014aapt}. Interviews were coded for general themes of faculty goals and the initial list included: developing skepticism in their own work, in science, and the media; understanding that measurements have uncertainty; developing agency over their own learning; communicating their results to a wider variety of audiences; learning how to use multiple sources of information to develop their understanding; demonstrating the ability to use and understand equipment; documenting their work effectively; and becoming reflective of their own experimental work.

With the intent of resolving the faculty's expressed goals with the AAPT Lab Guidelines, the goals were synthesized under larger headings, which aimed to combine and/or to connect seemingly disconnected goals. In addition, through a series of informational meetings that roughly 10-12 faculty attended regularly, how these goals were being combined and connected to interested faculty were reflected upon. Additional critiques and refinements of these goals were collected through notes taken during these meetings. Through several revisions, a set of four broad goals that faculty agreed reflected their views on the purpose of this part of laboratory courses was finalized. Additionally, these goals were also represented in the AAPT Lab Guidelines. The finalized goals are listed in Table \ref{tab:lgs} along with short description of each; they are enumerated (LG{\it X}) in order to refer to them in later sections.

\begin{figure*}
\includegraphics[clip, trim=15 100 15 100, width=0.8\linewidth]{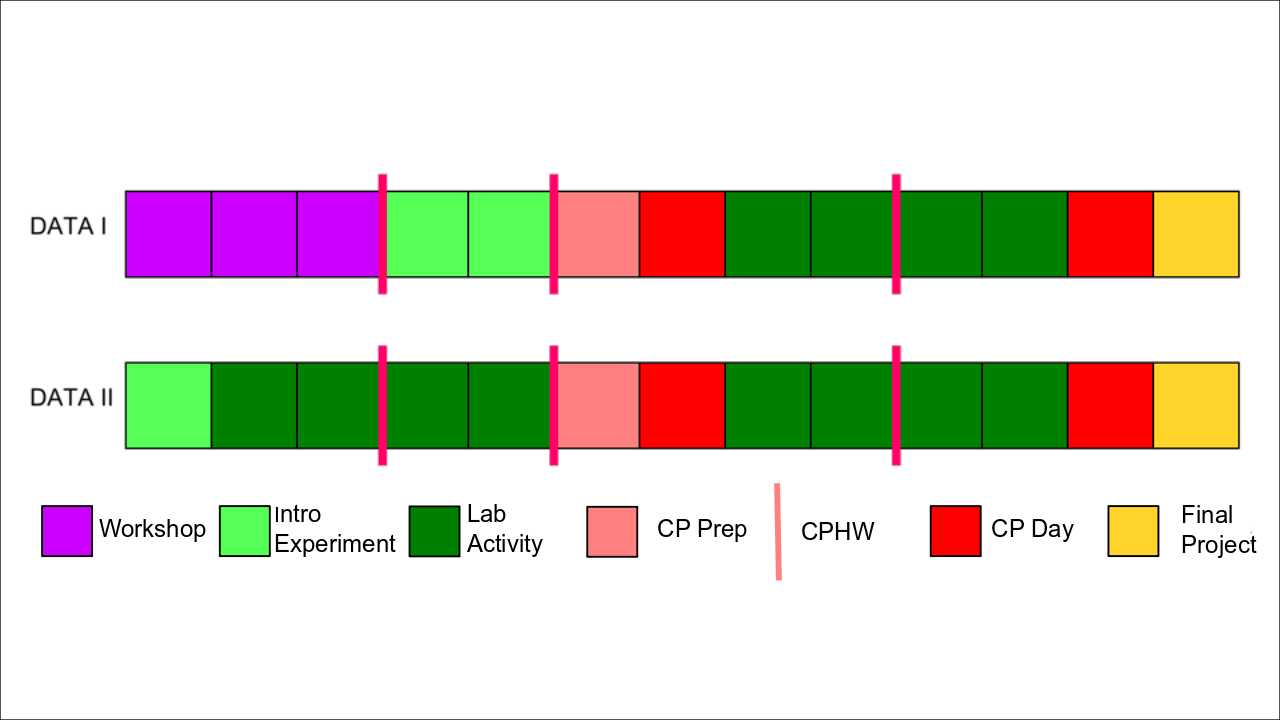}
\caption{Week-by-week schedule of DATA Lab I \& II.\label{weekly}}
\end{figure*}


The learning goals formed the basis for the design of course structures including materials and pedagogy. To construct these course structures, constructive alignment \cite{biggs1996enhancing} was leveraged, which helped ensure that the designed materials and enacted pedagogy were aligned with the overall learning goals for the course. These structures are described in the next section where we have included a direct reference to each learning goal that a particular course structure is supporting.

\section{Course Structures}
\label{structures}
Each laboratory section consists of twenty students and two instructors --  one graduate teaching assistant (GTA) and one undergraduate learning assistant (ULA) \cite{otero2010physics}. The students are separated into five groups of four, which they remain in for 4 to 6 weeks -- 4 to 6 class meetings. This time frame works well because it gives the students time to grow and improve as a group as well as individuals within a consistent group. In addition, when the groups are switched it requires the students to adapt to a new group of peers. The groups complete 6 (DL1) or 5 (DL2) experiments during the semester, most of them spanning two weeks -- two class meetings. Fig.~\ref{weekly} provides an overview of the two-semester sequence and will be unpacked further below. We indicate the laboratories that students complete with light green squares (introductory experiments) and dark green squares (two week labs). The students keep a written lab notebook, which they turn in to be graded at the end of each experiment.
\indent In this laboratory course, each group conducts a different experiment. This is possible because, in general, students tend to follow a similar path with respect to the learning goals and there is no set endpoint for any individual experiment. As long as students continue to work through the experimental process and complete analysis of their data, they are working towards the learning goals and can be evaluated using the aligned assessments (Sec.~\ref{assessments}). This approach also emphasizes that there is not one way to complete an experiment; this has added benefits for students' ownership and agency of the work as they must decide how to proceed through the experiment. In addition, having no set endpoint and two weeks to complete most experiments takes away the time pressure to reach a specific point in a given time. All of these aspects allow students to more fully engage with the work they are doing and, in turn, make progress toward the learning goals. Having each group conduct a different experiment addressed a significant point of discussion among the faculty; specifically, not covering the same breadth of content was a major concern. Although, through this design, students do not complete all of the experiments, they are introduced to all of the concepts through the peer evaluation of the communication projects (red squares in Fig.~\ref{weekly}, addressed in detail below).

\subsection{Laboratory Activities}

\begin{figure}
\includegraphics[clip, trim=110 375 100 180, width=0.8\linewidth]{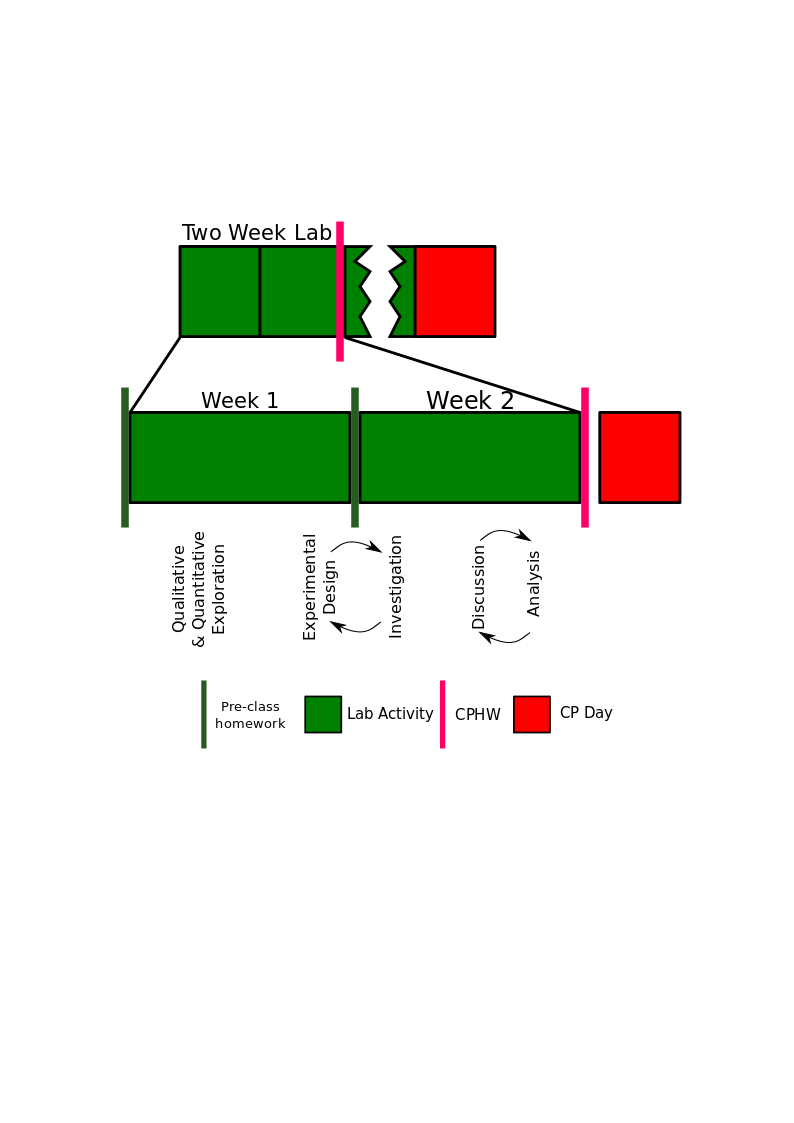}
\caption{A snapshot of an experiment from pre-class homework through the communication project.\label{snpsht}}
\end{figure}

The laboratory activities were designed around the learning goals. As such, the experiments follow a similar path from the beginning of the experimental process through analysis, with communication and collaboration as central components throughout. The course structures in relation to each of the learning goals are highlighted below. The core component (i.e. lab activities) of the course sequence is outlined in Fig.~\ref{snpsht}.\\
\textbf{LG1 - Experimental Process:} The students begin each experiment by broadly exploring the relevant parameters and their relationships. Typically, students investigate how changing one parameter affects another by making predictions and connecting their observations to physics ideas (qualitative exploration in Fig.~\ref{snpsht}). From these initial investigations, students work toward designing an experiment by determining what to measure, change, and keep the same. This often requires grounding decisions on some known model or an observed relationship (quantitative exploration, experimental design, and investigation in Fig.~\ref{snpsht}).\\
\textbf{LG2 - Data Analysis:} After additional formal investigations in which data has been collected, students summarize the raw data into an interpretable result. This typically includes some form of data analysis; for example, constructing a plot to evaluate a model or determining a quantitative relationship between the different variables in the data. In this work, the students are expected to make claims that are supported by their results. This often involves the students finding the slope and/or intercept in a plot and interpreting those results with respect to their expectations (discussion and analysis in Fig.~\ref{snpsht}).\\
\textbf{LG3 - Collaboration:} Throughout the experimental work and analysis, students discuss and make decisions with their peers in their lab group. Students are encouraged to develop a consensus approach to their work -- deciding collectively where to take their experiment and analysis. Furthermore, students are expected to make these decisions by grounding their discussions in their experiment, data, and analysis.\\
\textbf{LG4 - Communication:} Overall, the entire process requires that students communicate with their group and instructors. Additionally, students communicate their experimental approach and the results of their work including their analysis in their lab notebook. Later, students provide a more formal presentation of their work in the form of the communication projects.

It be should emphasized that this process is not content dependent; each laboratory activity conducted by a student group follows this process. This generalization enables the core components of the course to be repeated (see Fig. \ref{weekly}) to help address external constraints, such as limited equipment and time to work on experiments.

\subsection{Communication Projects}

DATA Lab is also defined by the focus on authentic scientific communication through the communication projects (CPs). The CPs are a formal way for the students to present their work and they are one of the assessments of the course in which the work done by the students is completed individually. CPs replace the lab practical from the traditional version of the course where students would conduct a smaller portion of a laboratory by themselves. CPs occur in the middle and at the end of the semester (red squares in Fig.\ref{weekly}). In DL1, the CP is a written proposal that summarizes the work the students conducted in one of their previous experiments and proposes an additional investigation. In DL2, the students create and present a research poster on one of (or a portion of one of) their experiments. In both courses, the projects are shared with and reviewed by their primary instructor and their peers in the class.

Through the CPs, students continue to engage with the faculty consensus learning goals (Sec.~\ref{LGSection}) as described below:\\
\textbf{LG1 - Experimental Process:} Students are expected to reflect on and summarize the process through which they went to complete the experiment. In so doing, they must communicate their rationale and reasoning for following that process.\\
\textbf{LG2 - Data Analysis:} The students must show that they can turn their raw data into an interpretable result. Again, this is often and, ideally, done in the form of a plot of their data with the emphasize of a model, including a fit, is needed. Students also present and explain what the results mean in the context of the experiment and a physical model.\\
\textbf{LG3 - Collaboration:} While the experiment was completed with the student's group where they may have consulted with their group mates, the CPs themselves are not inherently collaborative. However, in DL1, the reviews that students perform on each other's projects are done collaboratively in their groups.\\
\textbf{LG4 - Communication:} The CPs are the formal communication of a student's experimental work. In both courses, a student's CP is reviewed by their peers and feedback is provided describing successes and shortcomings along with suggestions for improvements.

\subsection{Final Projects}

The course structure was designed with the intent to provide students with a variety of ways to engage in the experimental physics practices. The final projects are an additional form of communication including an analysis and interpretation of experimental results through critiquing other scientific results (DL1--Critique Project) and describing a new experimental design (DL2--Design Project). 

\textit{Critique Project}: For the final project in DL1, students critique two sides of a popular science topic. In the prior week, students are arranged into new groups and before the class meeting, they must choose, as a group, from a list of possible topics such as climate change and alternative energy. In class, students collectively write up a summary and critique both sides of the scientific argument. 

\textit{Design Project}: For the final project in DL2, students choose an experiment that was conducted previously and design a new experiment for a future semester of DATA Lab. Similarly to DL1, the students are sorted into new groups and they must decide, as a group, which experiment they will be working on before the class meeting. Due to the structure of the course, specifically everyone doing different experiments throughout the semester, this choice may be an experiment that individual members of the group did not complete; negotiating this decision is part of the process of the Design Project. In class, students construct two documents: (1) a document that explains the design of the new experiment and (2) a document that would aide a future DATA Lab instructor to teach the experiment. Through this final project, DL2 students can design a project covering material that they may not have had the chance to explore during the course.

For both final projects, students turn in one assignment per group and they receive a single grade (as a group) for the assignment. Students also assess their own in-class participation, providing themselves a participation score (on a 4.0 scale) for the day. This score is submitted to their instructor along with their rationale for assigning themselves the grade.

These projects offer the final opportunity for DATA Lab students to engage with the faculty-consensus learning goals:\\
\textbf{LG1 - Experimental Process:} In DL1, students evaluate and summarize both sides of the chosen argument by reviewing the relevant data and experiments. Although students are not conducting an experiment, they are still asked to be critical of the experimental process in each side of the argument. In DL2, students must create a clear procedure for their proposed experiment. Here, they must consider the available equipment as well as how the data would be collected and why. \\
\textbf{LG2 - Data Analysis:} In DL1, the students must evaluate the evidence provided in each article. They must decide if there are obvious flaws in the way the analysis was conducted and if the analysis is compelling; that is, if the overall claims made in article align with the data and analysis. In DL2, students must consider the kind of analysis that would fit with their experiment and the data that they would collect. In addition, students are also expected to reflect on their analysis in light of the models that are available to explain the data they would collect. \\
\textbf{LG3 - Collaboration:} In both courses, students continue to work as a group and are graded accordingly. In addition, the students have been put into new groups, which they must adjust to.\\
\textbf{LG4 - Communication:} In both courses, students continue to communicate with their group as part of the collaboration.  In DL1 specifically, the final project provides an opportunity to communicate their own evaluation and critique of a scientific arguments. Students in DL2 are expected to communicate to different audiences, including future DATA Lab students and instructors, about their newly planned experiment.

\section{Overview of Key Supports}
\label{OverviewSection}	
As the students' work in this course is sufficiently open-ended, specific supports to ensure they feel capable of conducting the lab activities have been designed. Since the CPs are the main assessments in the DATA Lab course sequence and are a large portion of their overall grade for the course, the goals of the key supports are intended to provide students with the tools to help them succeed in the projects. Each of the supports designed for DATA Lab will be discussed in detail below (Secs.~\ref{sec:labs} \& \ref{sec:sup}). Assessments will be discussed in Sec.~ \ref{assessments}]

Broadly, the key supports for the students are outlined in Fig.~\ref{snpsht}. Before each class day, students complete a pre-class homework assignment (vertical green lines). Students also have three communication project homework (CPHW) assignments during the semester (vertical pink line) to help them complete their CPs. These supports, in addition to feedback on students' in-class participation and lab notebooks, apply for any of the regular two week experiments (green squares Fig.~\ref{weekly}). In the following section, these will be described in detail along with the additional supports that were designed for the courses.

\subsection{Typical Experiment}\label{sec:labs}
Each two-week experiment follows a similar path, highlighted in Fig.~\ref{snpsht} and described, in part, in Sec.~\ref{structures}. In this section, details of the general course components necessary to maintain the flexibility of the path students take through each experiment will be described.

\textbf{Pre-Class Homework:} At the beginning of an experiment, students are expected to complete the pre-class homework assignment which includes reading through the lab handout and investigating the suggested research. This assignment is usually 2-4 questions designed to have students prepare for the upcoming experiment. For example, before the first day of a new lab, students are asked what they learned during their pre-class research and if they have any questions or concerns about the lab handout. Between the first and second class meeting of the two-week experiment, students are expected to reflect on what they have already done and prepare for what they plan to do next. Typically, the 2-4 questions include reflections from the prior week, such as any issues their group ran into on the first day, and what they intend on doing during the second day of the experiment. Answers to the pre-class homework serve as additional information that the instructors can draw on during the class; knowing what questions and confusions that their students might have can help instructors be more responsive during class. Overall, the goal of the pre-class homework is for the students to come into class prepared to conduct their experiment and this assignment is used to hold them accountable for that preparation.  

\textbf{In-class Participation}: With the overall intent of improving students' specific laboratory skills and practices that are outlined in the course learning goals (Sec.~\ref{LGSection}), students receive in-class participation grades and feedback after every lab section (green squares in Figs~\ref{weekly} \&~\ref{snpsht}) on their engagement with respect to these practices. As the lab handouts do not provide students with specific steps that they must take to complete the experiment, students are expected to make most of the decisions together as a group. Generally, students have control over how their investigation proceeds; however, this control varies between experiments (i.e. students choose how to set up the equipment, what to measure, how to take measurements, etc.). The in-class participation grades and feedback are where students are assessed most frequently and where they have the quickest turnaround to implement the changes. See Sec.~\ref{sec:assA} for the details of how in-class participation is assessed. 

\textbf{Lab Notebooks}: For each experiment that the students engage in, they are expected to document their work in a lab notebook. In comparison to formal lab reports, lab notebooks are considered a more authentic approach to documenting experimental work. Furthermore, lab notebooks provide students with space to decide what is important and how to present it. The lab notebooks are the primary source that the students use to create their CPs. Like in-class participation, students receive lab notebook feedback much more regularly than CP feedback, so they have greater opportunity to reflect and make improvements. The specific details of the assessment of lab notebooks will be explained in Sec.~\ref{sec:assA}.

\textbf{CP Homeworks:} Three times during the semester the students complete CPHW assignments in addition to that week's pre-class homework. Each CPHW focuses on a relevant portion of the CPs (e.g., making a figure and a caption). Through the CPHWs, the aim is for students to develop experience with more of the CP components. In addition, students receive feedback on these different aspects (see Sec.~\ref{sec:assA}) , which they can act upon before they have to complete their final CPs. 

\textbf{Communication Projects:} Throughout each semester, the students complete two CPs, the first of which is a smaller portion of their overall course grade. With the goal of providing the students with a second opportunity to conduct a CP after receiving initial feedback, this course design feature intends to create less pressure on students during their first CP assignment. Students are expected to reflect on the process, their grade, and the feedback before they have to complete another CP. The CP assessment details will be discuss further in Sec.~\ref{sec:assB}.


\subsection{Additional Supports}\label{sec:sup}

Along with the support structures for the core components of the course sequence, additional supports have been designed to ease students into the more authentic features of DATA Lab such as designing experiments and documenting progress in lab notebooks. DL1 begins with three weeks of workshops (purple squares in Fig.~\ref{weekly}), followed by the introductory experiment (light green squares in Fig.~\ref{weekly}) that all of the students complete. DL2 begins with an introductory experiment as well, under the assumption that the students already went through DL1. The workshops and introductory experiments are designed to assist the students in navigating the different requirements and expectations of the overall course sequence, and of a typical experiment within each course. The additional support structures are described in detail below.

\textbf{DL1 Workshops:} The first workshop focuses on measurement and uncertainty with a push for the students to discuss and share their ideas (LG1,3). The students perform several different measurements -- length of a metal block, diameter of a bouncy ball, length of a string, mass of a weight, and the angle of a board. Each group discusses the potential uncertainty associated with one of the measurements. Then, students perform one additional measurement and assign uncertainty to it. The second workshop also focuses on uncertainty but in relation to data analysis and evaluating models (LG2,4) using the concept of a spring constant. Students collect the necessary measurements, while addressing the associated uncertainty and plot the measurements to analyze how the plot relates to the model of a spring. The final workshop focuses on proper documentation. The lab handouts do not contain their own procedure, so each student is expected to document the steps they take and their reasoning (LG4) in their lab notebook. In preparation for the third workshop, as a pre-class homework, students submit a procedure for making a peanut butter and jelly sandwich, which they discuss and evaluate in class. Students are then tasked with developing a procedure to determine the relationship between different parameters (length of a spring and mass added, angle of metal strip and the magnets placed on it, or time for a ball to roll down a chute and how many blocks are under the chute. At the end of each workshop the students turn in their notebooks, just as they would at the end of any experiment.

\textbf{Introductory Experiments:}  In DL1, the introductory experiment occurs after the three workshops. All students conduct a free-fall experiment where they must determine the acceleration due to gravity and the terminal velocity for a falling object. In DL2, the introductory experiment is the first activity in the course. This is because students will have already completed DL1 prior to taking DL2; rather than being slowly introduced to what DATA Lab focuses on, students can be reminded in a single experiment. The introductory experiment for DL2 involves Ohm's Law; students must determine the resistance of a given resistor. 

As these are the first DATA Lab experiments for either course, the instructors take a more hands-on and guiding approach than they will later in the semester. In DL1, these instructional changes represent a dramatic shift from the guidance students had during the workshops where instructors are often quite involved. In DL2, the one week lab is intended to be simple enough that students can be reminded of the expectations with respect to the overall learning goals of the course.

\textbf{CP Prep Day:} As discussed in the prior section, the CPs comprise a large portion of the students' total grade in the course. In addition to the supports that were already mentioned -- in class grades, notebooks, CPHW, and a lower stakes CP1 -- in the spring semester, the MSU academic calendar offers time for a communication project prep day (pink squares in Fig.~\ref{weekly}). This gives the students an extra day where they have time to work on their CPs in class. They can take additional measurements, seek help from their group or instructor, or work on the project itself. This prep day allows for a gentler transition into the CPs with a bit more guidance. It also reduces the amount of work that the students have to do outside of class.

\section{In Course Assessments}
\label{assessments}

The DATA Lab activities described above were designed around the overall learning goals outlined in Sec.~\ref{LGSection}. As such, the course assessments were also aligned with these overall course goals. There are two types of assessments used in DATA Lab -- formative (to help the students improve upon their work) and summative (to evaluate the students' output); these are separated for clarity. In this section, the various assessment tools are discussed with respect to the overall learning goals of the course.

\subsection{Formative Assessments} \label{sec:assA}
In DATA Lab the formative assessments are comprised of students' work on their in-class activities, lab notebooks, and CPHWs. Other than the pre-class homework, which is graded on completion, there is a rubric for each activity for which students receive a score. Each is structured to ensure that any improvements students make carry over to their CPs.

\textbf{In-class Participation}: In-class participation feedback is broken into group, which covers the general things everyone in the group or the group as a whole needs to work on, and individual, which is specific to the student and not seen by other group members. The general structure of the feedback follows an evaluation rubric used in other introductory courses and focuses on something they did well, something they need to work on, and advice on how to improve \cite{irving2017p3}. It is expected that students will work on the aspects mentioned in their prior week's feedback during the next week's class. Students are graded based on their response to that feedback. Any improvements they make with respect to the learning goals in class will also likely impact how well they complete their CPs.

Students' in-class participation is assessed with respect to two components, group function and experimental design. Specifically, group function covers their work in communication, collaboration, and discussion (LG3,4). For communication they are expected to contribute to and engage in group discussions. To do well in collaboration, students should come to class prepared and actively participate in the group’s activities. Discussion means working as a group to understand the results of their experiment. Experimental design evaluates the process that students take through the experiment and their engagement in experimental physics practices (LG1,2). They are expected to engage with and show competence in use of equipment, employ good experimental practices (i.e., work systematically, make predictions, record observations, and set goals) and take into account where uncertainty plays into the experimental process (i.e., reduce, record, and discuss it).

Specifically for the DL1 Workshops, instructors grade students differently than they would for a typical experiment. The emphasis for the workshops is on the group function aspect of the rubrics, communication and participation. This is because the students are being eased into the expectations that the instructors have around experimental work. 

\textbf{Lab Notebooks}: Feedback and grades for lab notebooks are only provided after the experiment is completed (the two week block in Figs~\ref{weekly} \&~\ref{snpsht}). Students receive individual feedback on their notebook, although members of a group may receive feedback on some of the same things simply because they conducted the experiment together. Like for in-class participation, it is expected that the students will work on the aspects mentioned in their feedback for the next lab notebook and the instructor can remind them of these things in class during the experiment.

Lab notebooks are also graded over two components, experimental design and discussion. Experimental design focuses on the experimental process and how students communicate it (LG1,4). Here, instructors typically look for clearly recorded steps and results, and intentional progression through the experiment. Discussion covers uncertainty in the measurements and the models, as well as the results, with respect to any plots and conclusions (LG2,4).  These evaluation rubrics for the lab notebooks were designed to be aligned with the those for the CPs, so that when students work toward improving their notebooks they are also making improvements that will benefit their CPs. For example, if a student is getting better at analyzing data and communicating their results within their notebooks, instructors should expect the same improvement to transfer to their CPs.\\
\indent For the DL1 Workshops, the lab notebooks are graded on the same components but the grades and feedback are specifically focused on the parts of the rubric that the students should have addressed in each of the previous workshops. For example, as documentation is emphasized in the last workshop, the students are not heavily penalized on poorly documented procedures in the first two workshops.\\
\textbf{CPHW}: The goal of this CPHW is to have students think about creating a more complete CP that connects their in-class work to the bigger picture. Students are evaluated on the quality and relevance of their sources, including the background and real-life connections (LG2,4). Each CPHW has a different rubric because each one addresses a different aspect of the CPs. 
\textit{Figure and caption}: The students create a figure with a robust caption based on the data from one of the labs they completed. Both the figure and the caption are evaluated on communication and uncertainty (LG2,4). For the plot, the students are expected to visualize the data clearly with error bars and it should provide insight into the various parameters within the experiment. For the caption, students need to discuss what is being plotted, make comparisons to the model including deviations, and draw conclusions that include uncertainty. \\
\textit{Abstract}: For a given experiment, students write a research abstract that covers the main sections of their project including introduction, methods, results, and conclusion. These are assessed on experimental process (motivation and clarity of the experiment) (LG1,4), and discussion (results and conclusions) (LG2,4).\\
\textit{Critique (DL1 only)}:  Students are given an example proposal that they must read, critique, and grade. This assignment plays two roles. First, students must examine a proposal, which should help to produce their own. Second, students must critique the proposal, which should help them provide better critiques to their peers. Students' performance is evaluated based on their identification of the different components of a proposal, and the quality of the feedback they provide (LG4).\\
\textit{Background (DL2 only)}: Students are tasked with finding three out-of-class sources related to one of their optics experiments, which they must summarize and connect back to the experiment.

\subsection{Summative Assessment} \label{sec:assB}
The CPs form the sole summative assessment of student learning in DATA Lab. As described above, each of the formative assessments are designed to align with the goals of the CPs.

\textbf{CPs}: As mentioned above, although students conduct the experiments together, the CPs are completed individually. In DL1, students' CP is a proposal that emphasizes their prior work and discusses a proposed piece of future work. As a result, the CP rubric is divided into two sections, prior and future work. Within those sections, there is a focus on experimental design and discussion. This rubric was iterated on after piloting the course for two semesters as it was found that students would often neglect either their future work or prior work when they were not directly addressed in the rubric; the rubrics were reorganized in order to account for this. Experimental design, which covers methods and uncertainty, focuses on the experimental methods and the uncertainty in measurements, models, and results when students discuss their prior work (LG1,2,4). In future work, experimental design refers to the proposed experimental methodology and the reasoning behind their choices (LG1,4). For the student's discussion of prior work, the rubric emphasizes how the they communicate their results (LG2,4). When students discuss their future work, the rubric emphasizes the novelty of the proposed experiment and the arguments made on the value of the project (LG1,4).

In DL2, students' CP is a poster that they present to their classmates for peer review. The rubric includes an additional component on the presentation itself, but the rubric still emphasizes the experimental design and discussion. Experimental design covers communication of the experimental process including students' reasoning and motivation. Discussion focuses on the discussion of uncertainty (i.e., in the measurements and models) and the discussion of results (i.e., in the plot and conclusions). The additional component focusing on presentation is divided into specifics about the poster (i.e., its structure, figures, layout) and the student's presentation of the project (i.e., clear flow of discussion, ability to answer questions).

\section{Example Experiment}
\label{ExpOverSection}
Overall, the course structures, supports, and assessments of DATA Lab have been discussed. In this section, the key supports will be grounded in examples from a specific experiment. The details of a specific two-week experiment will be described to better contextualize the features of the course. Additional experiments are listed in Tables \ref{DL1exp} and \ref{DL2exp} in the Appendix. The chosen experiment is from DL2 and is called ``Snell's Law: Rainbows''. In this experiment, students explore the index of refraction for different media and different wavelengths of light. 

Before attending the first day of the laboratory activity, students are expected to conduct the pre-class homework assignment, including the recommended research in Fig.~\ref{snells1}. In addition, the homework questions for the first day of a new experiment address the pre-class research, as follows:

\begin{figure}
\fbox{
\parbox{0.85\columnwidth}{
{\bf Research Concepts}\\

\begin{flushleft}
To do this lab, it will help to do some research on the concepts underlying the bending of light at interfaces including:
\begin{itemize}[noitemsep,nolistsep]
    \item Snell's Law (get more details than presented here)
    \item Refraction and how it differs from reflection
    \item Index of refraction of materials
    \item Fiber optics
    \item Using this simulation might be helpful: http://goo.gl/HEflDI
    \item How to obtain estimates for fits in your data (e.g., the LINEST function in Excel - http://goo.gl/wiZH3p)
\end{itemize}
\end{flushleft}

}}
\caption{Pre-class research prompts for the Snell's Law lab.\label{snells1}}
\end{figure}


\begin{figure*}[t]
\fbox{
\parbox{1.8\columnwidth}{
{\bf Part 1 - Observing Light in Water}\\

\begin{flushleft}
At your table, you have a tank of water and a green laser. Turn on the green laser and point it at the water's surface.
\begin{itemize}[noitemsep,nolistsep]
    \item What do you notice about the beam of light in the water?
    \item What about the path the light takes from the source to the bottom of the tank?
\end{itemize}
Let's get a little quantitative with this set up. Can you measure the index of refraction of the water? You have a whiteboard marker, a ruler, and a protractor to help you. Don't worry about making many measurements, just see if you can get a rough estimate by taking a single measurement.
\begin{itemize}
\item What does your setup and procedure look like for this experiment?
\item What part(s) of your setup/procedure is(are) the main source of uncertainty for this
measurement?
\item  Can you gain a sense of the uncertainty in this measurement?
\item  How close is your predicted value to the ``true value" of the index of refraction of mater?\\
\end{itemize}
\end{flushleft}

\vspace{11pt}
\noindent\rule{4in}{0.4pt}
\vspace{11pt}

\begin{flushleft}
On the optical rail you have a half circle shape of acrylic that is positioned on a rotating stage, with angular measurements. You also have a piece of paper with a grid attached to a black panel (i.e., a ``beam stop"). Using this setup, you will test Snell's Law for the green laser. Your group will need to decide how to set up your experiments and what measurements you will make. You should sketch the setup in your lab notebook and it would be good to be able to explain how your measurements relate to Snell's Law (i.e., how will the laser beam travel and be bent by the acrylic block?). In conducting this experiment, consider,
\begin{itemize}[noitemsep,nolistsep]
\item What measurements do you need to make?
\item What is the path of the laser beam and how does it correspond to measurements that you are making?
\item  What is a good experimental procedure for testing Snell's Law?
\item  What kind of plot is a useful one to convey how the model (Snell's Law) and your measurements match up?
\item Where is the greatest source of uncertainty in your experimental setup? What does that mean about the uncertainty in your measurements?
\end{itemize}
\end{flushleft}
}}
\caption{Snell's Law: Rainbows Lab Handout. \textit{Top}: Exploring refraction, first day of Snell's Law. \textit{Bottom}: Beginning model evaluation, main Snell's Law activity. \label{snells23}}
\end{figure*}

\begin{enumerate}
\item Describe something you found interesting in your pre-class research.
\item From reading your procedure, where do you think you may encounter challenges in this lab? What can you do to prepare for these?
\item Considering your assigned lab, is there anything specific about the lab handout that is unclear or confusing?
\end{enumerate}

The first day of the lab begins with exploring refraction in a water tank. Students are asked to qualitatively explore the index of refraction of the water using a simple setup (Fig.~\ref{snells23}). The exploration is fully student led; they investigate the laser and tank, discussing what they see with their group as they go and recording their observations in their notebooks. Students observe that the path of the light changes once the laser crosses the air-water boundary. Students are then lead to a quantitative exploration by determining the index of refraction of the water; instructors expect the students to have an idea of how to do this after their pre-class research. If students are not sure how to start, they are encouraged to search for Snell's Law online where they can quickly find a relevant example. The instructors check in with the students toward the end of this work. Typically, instructors will ask about the questions outlined in the lab handout.


The next part of the experiment is where students work to gain precision in their measurements and evaluate the model of the system. This part is most similar to a traditional laboratory course. The difference is that the students are told the goal but not how to proceed (see Fig.~\ref{snells23}). There are a number of decisions they must make as a group as they progress. Students record and explain their decisions in their lab notebooks; they might also discuss them with their instructor.


Typically by the end of the first day students know how to set up their experiment and have documented that in their lab notebooks. They are unlikely to have taken more than one measurement (the design and investigation phase in Fig.~\ref{snpsht}). They will return the following week to complete their experiment. The homework questions between the first week and the week that they return emphasize students' reflections on the previous week. Students also are asked think about the experiment outside of class. The typical homework questions prior to Week 2 are the following:

\begin{enumerate}
\item Because you will be working on the same lab this week, it is useful to be reflective on your current progress and plans. Describe where your group ended up in your current lab, and what you plan to do next.
\item Now that you are halfway through your current lab and are more familiar with the experiment, what have you done to prepare for this upcoming class?
\item Describe something that you found interesting in your current lab and what you would do to investigate it further.
\end{enumerate}

\begin{figure*}[t]
\includegraphics[clip, trim=0 0 0 0, width=0.8\linewidth]{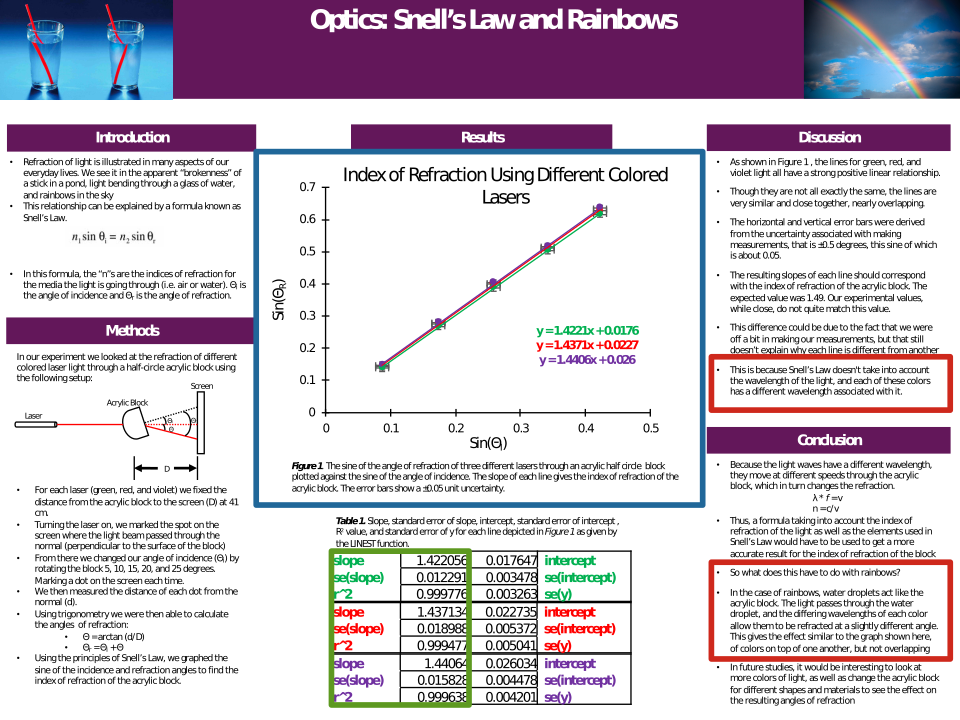}
\caption{Sample of a student's Communication Project for DL2. \textit{Blue}: graph with sine of the angle of incidence plotted against sine of the angle of refraction for each wavelength of light. \textit{Green}: The slope for each wavelength, which is the index of refraction of the block. \textit{Red}: Results and conclusions where they discuss the differences in the indices of refraction and how that is related to rainbows.\label{poster}}
\end{figure*}

The second week starts with setting up the experiment again and beginning the process of taking multiple measurements. At this point, students often break up into different roles: someone manipulating the equipment, one or two people taking measurements, and someone recording the data and/or doing calculations. These roles are what students appear to fall into naturally, and are not assigned to them. Although, if one student is always working in excel or always taking the measurements, instructors will address it in their feedback where they encourage the students to switch roles.

The next step depends on the amount of time that students have left in the class. If there is not much time, students focus on the data from one wavelength of light. If they have more time, they can make the same measurements with lasers of different wavelengths. In both cases, students can determine the index of refraction of the acrylic block. With multiple wavelengths, students are able to see that the index of refraction depends on wavelength. This leads to a conversation with the instructor about how this relates to rainbows and a critique of the model of refraction -- Snell's Law.

Most of the analysis that students conduct in this example experiment is the same regardless of how many lasers they collected data (discussion and analysis in Fig.~\ref{snpsht}). While considering the different variables in their experiment, students are expected to make a plot where the slope tells them something about the physical system. In this case, the design is intended for the students to plot the sine of the angle of incidence on the x-axis and the sine of the angle of refraction on the y-axis, which makes the slope the index of refraction of the acrylic block. The optics experiments occur in the second half of the semester after the students have become familiar with constructing linear plots from nonlinear functions. For this lab, students usually do not have much difficulty determining what they should plot. After they obtain the slope and the error in the slope, students will typically compare it to the known index of refraction of the acrylic block. They must research this online as it is not provided anywhere for them in the lab handout.

The second day of the experiment ends with a discussion of their plot. Students construct a conclusion in their notebooks that summarizes the results, what they found, what they expected, reasons for any differences, and an explanation of what it all means in the larger physics context.

After the experiment, the students may have their third and final CPHW, background/literature review. In the case of Snell's Law, students would be asked to find three additional sources where these concepts are used in some other form of research, often in the field of medicine but also in physics or other sciences. Students then summarize what they did in class and connect their experimental work to the sources that they found.

The student can choose to do their second CP on this experiment. An example of a poster can be seen in Fig.~\ref{poster}. In the figure, three key features are highlighted. First, in the blue box, is the graph where students plotted all three wavelengths of light. In the green box, is the slope for each color, which is the index of refraction of the acrylic for each laser. Finally, in the red boxes, are their results and conclusion. In the top box, students explained why their indices are different, that is, because of the assumption that Snell's Law is wavelength independent. In the bottom box, they make the connection to rainbows. The student would present this poster during the in-class poster session, to their peers and their instructor.


\section{Redesign Efficacy}
\label{efficacy}

To measure the efficacy of the DATA Lab course transformation, the Colorado Learning Attitudes about Science Survey for Experimental Physics (E-CLASS) \cite{zwickl2014epistemology} was implemented in the traditional laboratory course as well as the transformed courses. The E-CLASS is a research-based assessment tool used to measure students' epistemology and expectations about experimental physics \cite{wilcox2016students,hu2017qualitative,wilcox2018summary}. The well-validated survey consists of 30 items (5-point Likert scale) where students are asked to rate their level of agreement with each statement. The scoring method of this assessment was adapted from previous studies \cite{adams2006new}. First, the 5-point Likert scale is compressed into a 3-point scale; ``(dis)agree" and ``strongly (dis)agree" are combined into one category. Then, student responses are compared to the expert-like response; a response that is aligned with the expert-like view is assigned a $+1$ and a response that is opposite to the expert-like view is assigned a $-1$. All neutral responses are assigned a 0. For our comparison between the traditional and transformed courses, we will report the percentage of students with expert-like responses.

\begin{figure}
\includegraphics[clip, trim=0 0 0 0, width=\linewidth]{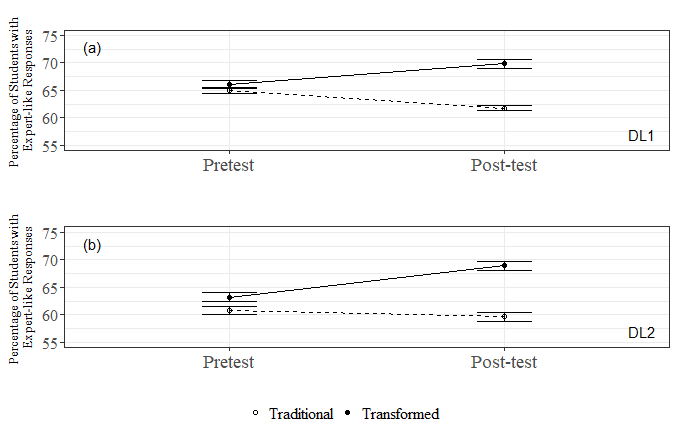}
\caption{Fraction of students with expert-like responses for (a) DL1 and (b) DL2.\label{eclass}}
\end{figure}

In DL1 and DL2, the E-CLASS was administered as an online extra credit assignment both pre- and post-instruction. Throughout the course transformation, DL1 and DL2 collected a total of 1,377 and 925 students, respectively, with matched (both pretest and post-test) E-CLASS scores. Figure \ref{eclass} shows the fraction of students with expert-like responses in the traditional course and the transformed course for (a) DL1 and (b) DL2. Students in the traditional courses had a decrease of 3\% and and 1\%, respectively, in their expert-like attitudes and beliefs toward experimental physics from pre- to post-instruction. However, in the transformed DATA Lab courses, the students' expert-like views of experimental physics increased by 4\% in DL1 and by 6\% in DL2.
To explore the impact of the course transformation after controlling for students' incoming epistemology and expectations about experimental physics, ANCOVA was used to evaluate the student's attitudes and beliefs post-instruction between the traditional courses and the transformed courses. For both DL1 and DL2, results showed that there was a significant difference in ECLASS post-test percentages between the traditional courses and the transformed courses ($ps<0.001$). Specifically for DL1, results demonstrated a significant 7\% post-test difference in expert-like responses between the traditional course and the transformed course after controlling for ECLASS pretest scores. For DL2, there was a significant 9\% difference in post-test responses between the traditional and transformed courses after controlling for the student's incoming ECLASS responses. Overall, the transformation in both DL1 and DL2 had a positive impact on students' epistemological views and expectations about experimental physics.

\section{Conclusion}
\label{conclusions}

In this paper, the large scale transformation of the MSU algebra-based physics labs for life science students was described. The design was divorced from the specific physics content because the learning goals developed from a faculty consensus design did not include specific content. This design means that the individual lab activities do not matter {\it per se}, but instead the structure of the course and how students work through the lab are what is important. Theoretically, one could adapt this design to a chemistry or biology lab by making adjustments to the kinds of lab activities, and relevant changes to the learning goals. That being said, there are still key structures to ensure the functioning of the course which will be covered in detail in a subsequent paper (e.g. a leadership team of four instructors, two GTAs and two ULAs, tasked with maintaining consistent grading and instruction across the sections).


The transformation was centered to emphasize experimental physics practices. The overall efforts were focused on the two course series because the majority of the students that are taking courses in the physics department at MSU are enrolled in the introductory algebra based series, specifically 2,000 students per year. In addition, the majority of the student instructors in the MSU physics and astronomy department, nearly 80 graduate teaching assistants and undergraduate learning assistants, teach in these labs. Because of its scale, special attention was given to the voice of the physics faculty in the development of the learning goals for DATA Lab \cite{wieman2017improving}. The entire course was designed around the faculty-consensus learning goals, which are all based around physics laboratory practices (Sec.~\ref{LGSection}). From course structures to assessments, everything was intentionally aligned with the overall learning goals. Each component of the course builds upon another through the two semester sequence. Each individual lab activity builds upon skills that will be valuable for each subsequent activity, from lab handouts to pre-class homework assignments. Such an effort was put into designing this course sequence in large part because of the number of MSU undergraduate students they are serving. The value in physics labs for these non-majors lies in the scientific practices on which the redesign was centered. Those skills and practices are what they will take with them into their future careers.

\begin{acknowledgments}

This work was generously supported by the Howard Hughes Medical Institute, Michigan State University's College of Natural Science, as well as the Department of Physics and Astronomy. The authors would like to thank the faculty who participated in the discussion of learning goals. Additionally, we would like to thank S. Beceiro-Novo, A. Nair, M. Olsen,  K. Tollefson, S. Tessmer, J. Micallef, V. Sawtelle, P. Irving, K. Mahn, J. Huston who have supported the development and operation of DATA Lab. We also thank the members of the Physics Education Research Lab who have given feedback on this work and this manuscript.

\end{acknowledgments}

\bibliography{data-lab}
\bibliographystyle{apsper}

\appendix*   

\section{Experimental Descriptions}
\begin{table}[ht!]
\caption {DL1 Experiments}
\begin{ruledtabular}
\begin{tabular}{l p{5cm}}
Experiment & Main Idea \\
\hline	
Workshop 1 &  Taking a variety of measurements and determining the uncertainty \\
Workshop 2 & Taking simple measurements, learning how to plot parameters in useful ways, and incorporating uncertainty into analysis and results \\
Workshop 3 & Using the example of making a peanut butter and jelly sandwich to evaluate procedures and discuss proper documentation \\
\hline
Free Fall &  Exploring free fall and terminal velocity with coffee filters \\
\hline
(In)elastic Collisions & Exploring elastic and inelastic collisions on an air track \\
Standing Waves & Exploring waves on a string, nodes, and resonant frequencies \\
Impulse & Exploring impulse with bouncing balls and relating that to concussions \\
Energy Transfer & Exploring energy loss of a bouncing ball \\
Pendulum & Exploring a simple pendulum \\
\end{tabular}
\end{ruledtabular}
\label{DL1exp}
\end{table}



\begin{table}[ht!]
\caption {DL2 Experiments}
\begin{ruledtabular}
\begin{tabular}{l p{5cm}}
Experiment & Main Idea \\
\hline	
Ohm's Law &  Measuring current and voltage to determine resistance of a resistor \\
\hline
Fields & Exploring electric fields, determining the number of electrons transferred between two pieces of tape \\
Transformers & Exploring solenoids and transformers, developing a formula to describe the relationship between voltage and number of turns \\
Resistivity &  Exploring how resistance depends on length and width, determining the resistivity of clay \\
Charge-to-mass Ratio & Finding the charge-to-mass ratio of an electron \\
Biomeasures & Exploring an oscilloscope, measuring a heartbeat on the scope \\
\hline
Snell's Law: Rainbows & Exploring refraction of light, investigating Snell's Law and different wavelengths of light \\
Snell's Law: Fiber Optics & Exploring refraction of light, investigating Snell's Law and total internal reflection \\
Lenses & Exploring thin lenses, investigating how corrective lenses work with different eye impairments \\
Interference & Exploring interference and diffraction, investigating single and double slit diffraction \\
Polarization & Exploring polarizers, investigating polarization to come up with Malus's Law \\
\end{tabular}
\end{ruledtabular}
\label{DL2exp}
\end{table}

\end{document}